\documentclass{ws-procs961x669}            
\usepackage[caption=false]{subfig}

\begin{document}
\title{Searching For Strange Quark Planets}

\author{Xu Wang$^{1,*}$, Yong-Feng Huang$^{1,2,*}$ and Bing Li$^{3,4}$}

\address{$^{1}$School of Astronomy and Space Science, Nanjing University,
	     Nanjing 210023, People's Republic of China  \\
	$^{2}$Key Laboratory of Modern Astronomy and Astrophysics (Nanjing University),
	     Ministry of Education, People's Republic of China  \\
    $^{3}$Key Laboratory of Particle Astrophysics, Institute of High Energy Physics,
         Chinese Academy of Sciences, Beijing 100049, People's Republic of China \\
    $^{4}$Particle Astrophysics Division, Institute of High Energy Physics,
         Chinese Academy of Sciences, Beijing 100049, People's Republic of China\\
	$^*$E-mail: wx\_tba@163.com; hyf@nju.edu.cn
}

\begin{abstract}
	 Strange quark matter (SQM) may be the true ground state of matter. According to
     this SQM hypothesis, the observed neutron stars actually should all be
	 strange quark stars. But distinguishing between neutron stars and strange quark stars
	 by means of observations is extremely difficult. It is interesting to note that
     under the SQM hypothesis, less massive objects such as strange quark planets and
     strange dwarfs can also stably exist. The extremely high density
	 and small radius of strange quark planets give us some new perspectives to
     identify SQM objects and to test the SQM hypothesis. First, the tidal
     disruption radius of strange quark planets is much smaller than normal planets,
     so, very close-in exoplanets can be safely identified as candidates of SQM
	 objects. Second, gravitational waves (GW) from mergers of strange quark star-strange quark
	 planet systems are strong enough to be detected by ground-based GW detectors. As a
	 result, GW observation will be a powerful tool to probe SQM stars. At the same time,
     the tidal deformability of SQM planets can be measured to further strengthen the result.
\end{abstract}


\section{Introduction} \label{aba:sec1}

The central engine of gamma-ray bursts may be compact objects such as neutron stars.
However, our knowledge about matter under extreme densities is still quite poor so
that the internal composition and structure of neutron stars are largely uncertain to
us \citep{2004Sci...304..536L}. It has been argued that the energy per baryon of
strange quark matter (SQM), which is composed of three favors of quarks (up,
down and strange quarks), could be less than normal hadronic matter. As a
result, SQM may be the true ground state of
matter \citep{1970PThPh..44..291I,1971ApJ...170..299B}. If this hypothesis
is true, then it is even possible that all the so called ``neutron stars''
observed in the Universe should actually be strange quark
stars\citep{1984PhRvD..30.2379F,1984PhRvD..30..272W}.

However, it is difficult to distinguish between strange quark stars and neutron stars through
current astronomical observations. For a long time, people have tried to find the difference
between neutron stars and strange quark stars in terms of mass-radius relationship,
cooling rate, the minimum spin period and gravitational wave (GW) features. But for a typical
1.4 M$_\odot$ compact object, the radius difference between a neutron star and a quark star
is too small to be detected through current observational technology.
Recently, Geng et al pointed out that fast radio busts may originate from the collapses of the
crust of strange quark stars \citep{2021.200152G,2018ApJ...858...88Z}. It provides an novel
visual angle on these interesting objects.

It is interesting to note that, under the SQM hypothesis, quark matter is bounded by strong
interaction but not gravity\citep{1984PhRvD..30..272W}. So, SQM can even exist stably in
the form of small chunks in the universe. It implies that planets composed of
strange quark matter can also exist stably\citep{1995PhRvL..74.3519G,1995ApJ...450..253G}.
Strange quark planets are very different from normal planets. They have a much higher mean
density and a much smaller radius, which provide us with some effective new methods to
test the SQM hypothesis. In our previous studies, we have suggested some new methods to
identify strange quark
planet \citep{2017ApJ...848..115H, 2020ApJ...890...41K, 2015ApJ...804...21G, 2021arXiv210513899W, 2019AIPC.2127b0027K, 2020arXiv201205748K}.
In this article, we will summarize the results.

\section{Searching for strange quark planets among close-in exoplanets} \label{aba:sec2}

Planets cannot be too close to their host stars, otherwise they will be tidally
broken up by the strong tidal force of their hosts. We can use the tidal disruption
radius $r_{td}$ to describe the shortest possible separation between a planet and its host,
which can be analytically expressed as\citep{1975Natur.254..295H}
\begin{equation}
    r_{td}\thickapprox\left(\frac{6M}{\pi \bar{\rho}}\right)^{\frac{1}{3}},
\end{equation}
where $M$ is the mass of the host star and $\bar{\rho}$ is the mean density of the planet.
For the convenience of calculation, this equation can be further written as,
\begin{equation}
    r_{td}\thickapprox2.37\times10^6\left(\frac{M}{1.4\rm{M_{\odot}}}\right)^\frac{1}{3}\times\left(\frac{\bar{\rho}}{4\times10^{14}\rm{g cm^{-3}}}\right)^{-\frac{1}{3}} \rm{cm}.
\end{equation}

Suppose a planet orbits around a host star which has a typical mass of 1.4 $\rm{M_{\odot}}$.
For a strange quark planet, its mean density can be as high as $4\times10^{14}\ \rm g\,cm^{-3}$.
With such a high density, the $r_{td}$ of the strange quark planet will be less
than $2.37\times10^{6}\ \rm cm$, which is only about twice the
radius of a pulsar\citep{2017ApJ...848..115H,2020ApJ...890...41K}. But for normal matter
planets, the density is on the order of $\sim 8\ \rm g\,cm^{-3}$, and the corresponding
$r_{td}$ is generally larger than $8.7\times10^{10}\ \rm cm$. Even if we take the
mean density as $30\ \rm g\,cm^{-3}$, which is already a very high
value for normal matter, the derived $r_{td}$ will still be larger than $5.6\times10^{10}\ \rm
cm$\citep{2017ApJ...848..115H,2020ApJ...890...41K}. From these simple calculations, we
argued that if the orbital radius of a planet is found to be less than $5.6\times10^{10}\ \rm cm$,
then the planet should be a strange quark planet.

An extremely close-in strange quark planet cannot be observed directly by imaging method.
On the contrary, it can be relatively easily detected via pulsar timing observations.
According to Kepler's law, the relationship between the orbital radius and the
period can be expressed as
\begin{equation}
    \frac{a^3}{P^2_{\rm orb}}\thickapprox\frac{GM}{4\pi^2},
\end{equation}
where $G$ is the gravity constant, $a$ is the orbital radius and $P_{\rm orb}$ is
the planet's orbital period. For planets with $a$ smaller than $\sim 5.6\times10^{10}\ \rm cm$,
the orbital period will be less than $\sim 6100$ s. Therefore, it is argued a planet with the
orbital period less than 6100 s should be a strange quark
planet\citep{2017ApJ...848..115H,2020ApJ...890...41K}.

Using the above criterion, Kuerban et al. have tried to search for strange quark planets
among exoplanets, especially among pulsar planets \citep{2020arXiv201205748K}.
According to their results, the short period pulsar planets of PSR J0636 b,
PSR J1807-2459A b and PSR 1719-14 b are good candidates of strange quark planets.

\section{Searching for strange quark planets through GW observations} \label{aba:sec3}

Since 2015, gravitational wave observations have opened a new window for
astronomy\citep{ 2016PhRvL.116v1101A} and are also expected to be a new tool for searching for
quark stars\citep{2010PhRvD..82b4016P}. Note that it is still quite difficult to distinguish between
binary neutron star mergers and binary quark star mergers by gravitational wave observations, because
these two kinds of compact stars have marginal difference in radius at the typical mass of 1.4
M$_\odot$ \citep{2010PhRvD..81b4012B,2014MNRAS.445L..11M}. However, gravitational waves may bring
new opportunities in searching for strange quark planets.

A normal matter planet can not be too close to its host, otherwise it will be tidally
disrupted. Consequently, the gravitational wave emissions from normal planetary systems
are usually too weak to be detected. But due to the extremely high density and the
very close-in orbit of strange quark planets, a strange star-strange planet system can
produce very strong gravitational wave emissions, especially at the final stage of the
merging process. If such a merger event occurs in the local universe, it would be
detectable for the gravitational wave detectors such as Advanced LIGO and the Einstein
Telescope \citep{2015ApJ...804...21G}.

GW observations can also help diagnose the internal composition and internal structure of
compact stars by means of tidal deformability measurements. The first tidal deformability
measurement has been obtained for the binary neutron star merger event of GW170817,
which gives a new constraint on the equation of state of dense
matter\citep{2017PhRvL.119p1101A,2019JPhG...46l3002G}.

Tidal deformability is a quantity that describes the deformation of a star in a tidal field,
which is defined as
\begin{equation}
\lambda=-\frac{Q_{ij}}{E_{ij}},
\end{equation}
where $Q_{ij}$ is the induced quadrapole moment of the star and $E_{ij}$ is the
tidal field that it resides in (i.e., the gravity field produced by its companion).
For convenience, this quality is often written in a dimensionless form, i.e.,
\begin{equation}
\Lambda=\frac{\lambda c^{10}}{G^4m^5}.
\end{equation}
Here, $c$ is the speed of light in vacuum and $m$ is the mass of the star.

Generally speaking, a larger tidal deformability means that the star will be relatively
easier to be deformed in a tidal field\citep{2019JPhG...46l3002G}. So, tidal deformability
will affect the evolution of the gravitational wave phase, which could be perceived in
GW observations \citep{2021arXiv210513899W}. Comparing the observed tidal deformability
with that calculated by solving the Tolman-Oppenheimer-Volkoff equation adopting a
particular equation of state (EoS), we can get useful information on the internal structure
of compact stars \citep{2008ApJ...677.1216H,2010PhRvD..81l3016H}.

\begin{figure}[htbp]
 \centering
 \subfloat{
  \includegraphics[width=0.75\linewidth]{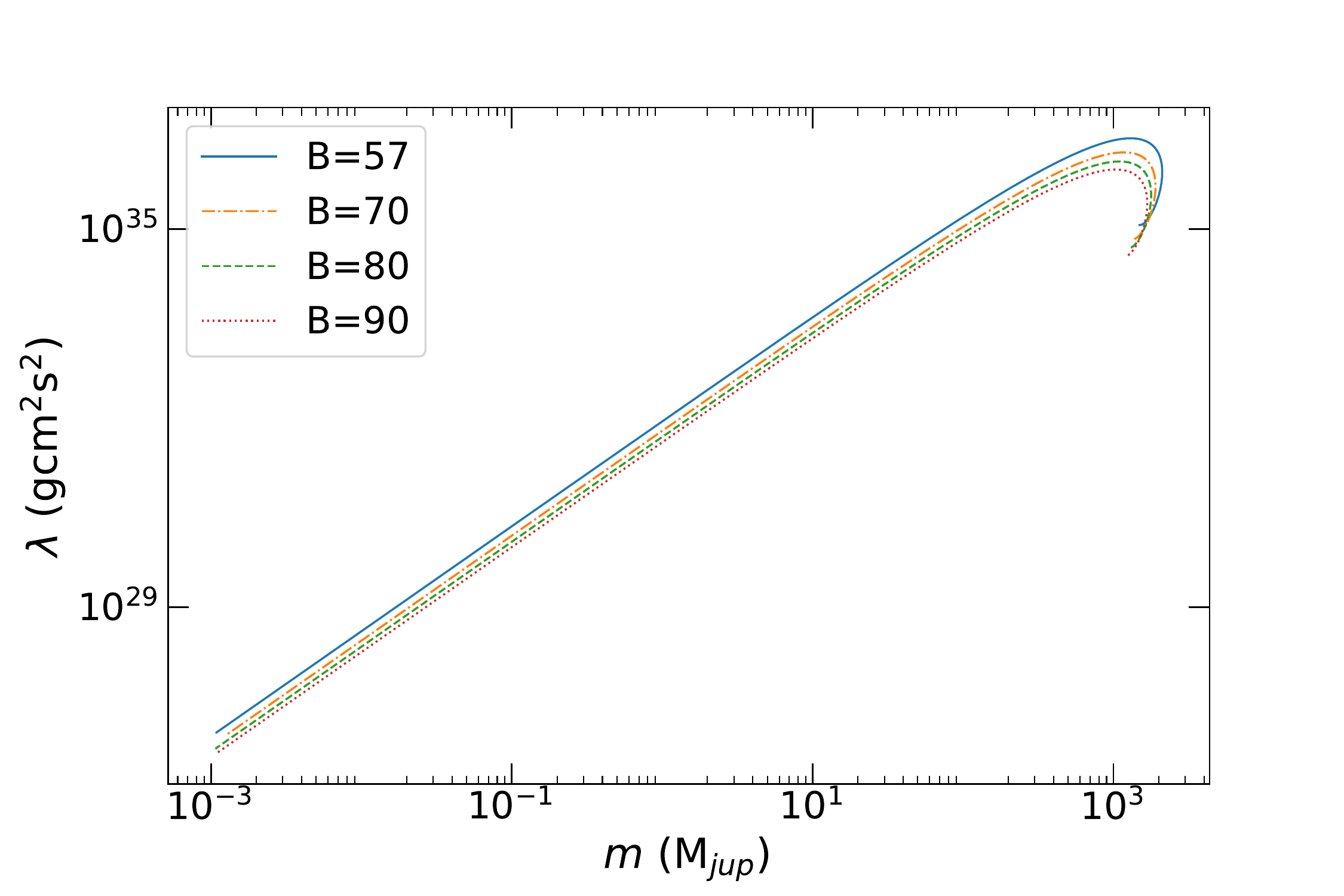}
 } \\
 \subfloat{
  \includegraphics[width=0.75\linewidth]{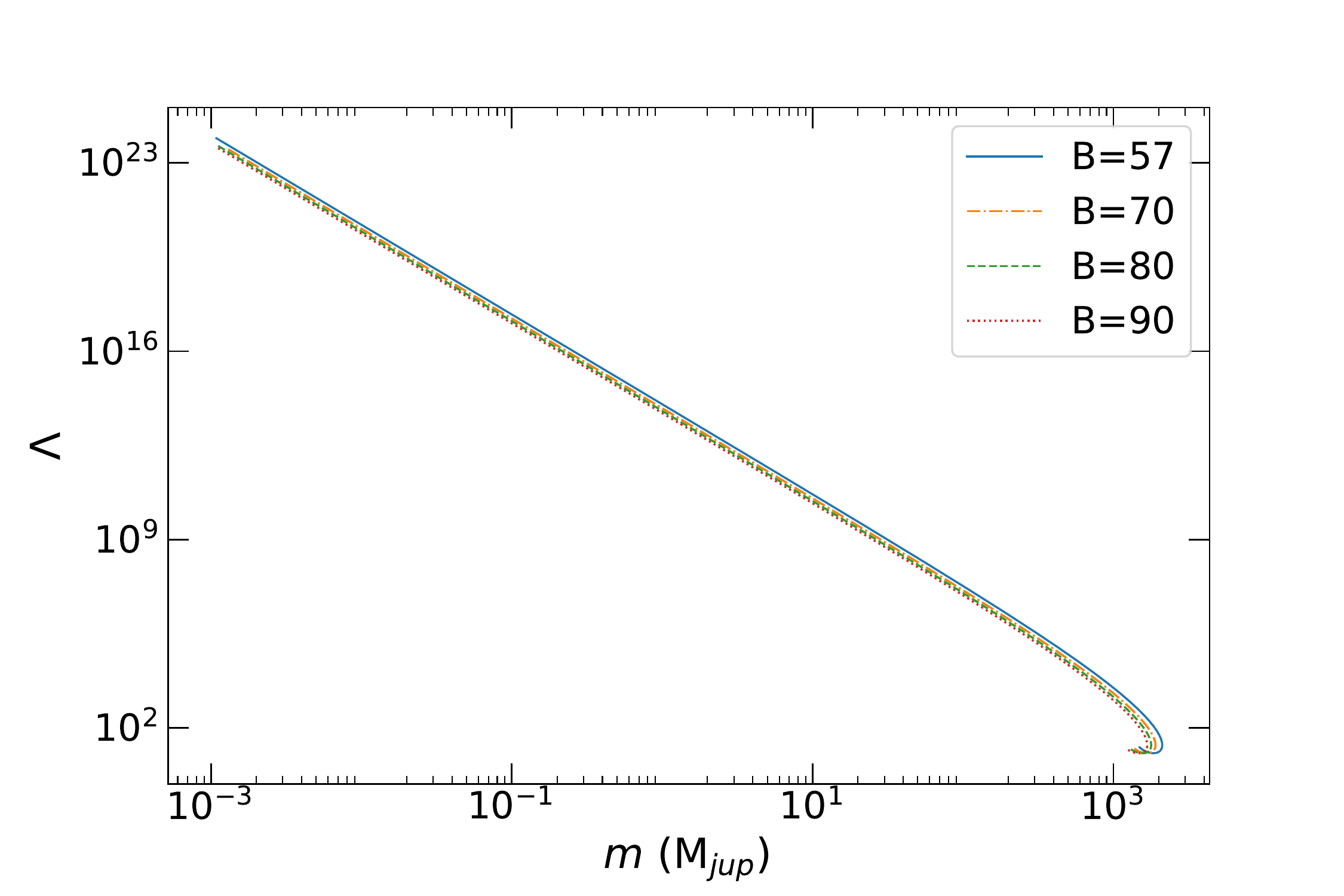}
 }
 \caption{Tidal deformability (the upper panel) and dimensionless tidal deformability (the lower
    panel) versus mass for strange quark stars and strange quark planets\citep{2021arXiv210513899W}.
    Different line styles represent different bag constant, which is
    marked in the figure in units of $ \rm MeV/fm^3$.   }
 \label{fig:s}
\end{figure}

Following the method of Hinderer et al.\citep{2008ApJ...677.1216H,2010PhRvD..81l3016H},
we have calculated the tidal deformability of strange quark stars, paying special
attention on strange dwarfs and strange planets. We engage the bag model for strange
quark matter\citep{1984PhRvD..30.2379F}. Our results are shown in Figure \ref{fig:s}.
We see that as the mass decreases, the value of the deformability also decreases,
while the value of the dimensionless tidal deformability keeps increasing. For
a planet with a mass of $10^{-3} M\rm_{jup}$, its tidal deformability is up to $\sim 10^{27}$
g cm$^2$ s$^2$ and its dimensionless tidal deformability is up to $\sim 10^{23}$,
which are much higher than those of normal matter planets. Therefore, the
tidal deformability is a useful parameter for identifying strange quark planets\citep{2021arXiv210513899W}.

\section{Summary} \label{aba:sec4}

Discriminating between neutron stars and strange quark stars is a challenging task.
In this article, we try to solve the problem from a novel point of view. We mainly
concentrate on strange quark planets. It is suggested that there are basically two
methods to search for strange quark planets. First, we can try to find close-in
pulsar planets with the orbital period less than 6100 s. Encouragingly, at least
three possible strange quark planet candidates have been found. Second, we can identify
strange quark planets through gravitational wave observations. It is found that the mergers
of strange star-strange planet systems can produce strong gravitational wave bursts,
which can potentially be detected by the advanced LIGO and the future Einstein Telescope.
The tidal deformability of strange quark dwarfs and strange quark planets is specially
calculated. In short, we stress that strange quark planets could be a powerful tool for
clarifying the nature of the so called ``neutron stars''.

\section*{Acknowledgements}
This work is supported by National SKA Program of China No. 2020SKA0120300, by the National
Natural Science Foundation of China (Grant Nos. 11873030, 12041306, U1938201), and by the
science research grants from the China Manned Space Project with NO.CMS-CSST-2021-B11.

\bodymatter

\bibliographystyle{ws-procs961x669}
\bibliography{MS}

\end{document}